\documentclass[twocolumn,floatfix,showpacs,prb]{revtex4}
\pdfoutput=1
\usepackage{graphicx}
\usepackage{amsmath}
\usepackage{amssymb}
\usepackage{bm}

\begin{document}

\title{Andreev reflection from a topological superconductor with chiral symmetry}
\author{M. Diez}
\affiliation{Instituut-Lorentz, Universiteit Leiden, P.O. Box 9506, 2300 RA Leiden, The Netherlands}
\author{J. P. Dahlhaus}
\affiliation{Instituut-Lorentz, Universiteit Leiden, P.O. Box 9506, 2300 RA Leiden, The Netherlands}
\author{M. Wimmer}
\affiliation{Instituut-Lorentz, Universiteit Leiden, P.O. Box 9506, 2300 RA Leiden, The Netherlands}
\author{C. W. J. Beenakker}
\affiliation{Instituut-Lorentz, Universiteit Leiden, P.O. Box 9506, 2300 RA Leiden, The Netherlands}
\date{June 2012}
\begin{abstract}
It was pointed out by Tewari and Sau that chiral symmetry ($H\mapsto -H$ if $\text{e}\leftrightarrow\text{h}$) of the Hamiltonian of electron-hole (e--h) excitations in an $N$-mode superconducting wire is associated with a topological quantum number $Q\in\mathbb{Z}$ (symmetry class BDI). Here we show that $Q={\rm Tr}\,r_{\rm he}$ equals the trace of the matrix of Andreev reflection amplitudes, providing a link with the electrical conductance $G$. We derive $G=(2e^{2}/h)|Q|$ for $|Q|=N,N-1$, and more generally provide a $Q$-dependent upper and lower bound on $G$. We calculate the probability distribution $P(G)$ for chaotic scattering, in the circular ensemble of random-matrix theory, to obtain the $Q$-dependence of weak localization and mesoscopic conductance fluctuations. We investigate the effects of chiral symmetry breaking by spin-orbit coupling of the transverse momentum (causing a class BDI-to-D crossover), in a model of a disordered semiconductor nanowire with induced superconductivity. For wire widths less than the spin-orbit coupling length, the conductance as a function of chemical potential can show a sequence of $2e^{2}/h$ steps --- insensitive to disorder. 
\end{abstract}
\maketitle

\section{Introduction}

The classification of topological states of matter, the socalled ``ten-fold way'',\cite{Alt97} has five topologically nontrivial symmetry classes in each dimensionality.\cite{Ryu10} For a one-dimensional wire geometry, two of these five describe a topological superconductor and the other three a topological insulator. Each symmetry class has a topological quantum number $Q$ that counts the number of protected bound states at the end of the wire. These end states are of particular interest in the topological superconductors, because they are pinned at zero excitation energy by electron-hole symmetry and are a condensed matter realization of Majorana fermions.\cite{reviews} Signatures of Majorana zero-modes have been reported in conductance measurements on InSb and InAs nanowires, deposited on a superconducting substrate.\cite{Mou12,Den12,Das12}

A key distinction between superconducting and insulating wires is that $Q\in\mathbb{Z}_{2}$ is a parity index in a topological superconductor, while all integer values $Q\in\mathbb{Z}$ can appear in a topological insulator. In other words, while there can be any number of protected end states in an insulating wire, pairs of Majorana zero-modes have no topological protection. The symmetry that prevents the pairwise annihilation of end states in an insulating wire is a socalled chiral symmetry of the Hamiltonian: $H\mapsto-H$ upon exchange $\alpha\leftrightarrow\beta$ of an internal degree of freedom, typically a sublattice index.

In an interesting recent development,\cite{Tew11} Tewari and Sau have argued (motivated by Ref.\ \onlinecite{Niu12}) that an approximate chiral symmetry may stabilize pairs of Majorana zero-modes in a sufficiently narrow nanowire. The symmetry $H\mapsto-H$ when $\text{e}\leftrightarrow\text{h}$ involves the exchange of electron and hole indices e,h.\cite{Che10} It is distinct from electron-hole symmetry, which involves a complex conjugation $H\mapsto-H^{\ast}$ and is a fundamental symmetry of the problem. The combination of chiral symmetry and electron-hole symmetry promotes the superconductor from symmetry class D to symmetry class BDI, extending the range of allowed values of $Q$ from $\mathbb{Z}_{2}$ to $\mathbb{Z}$.

\begin{figure}[tb] 
\centerline{\includegraphics[width=0.8\linewidth]{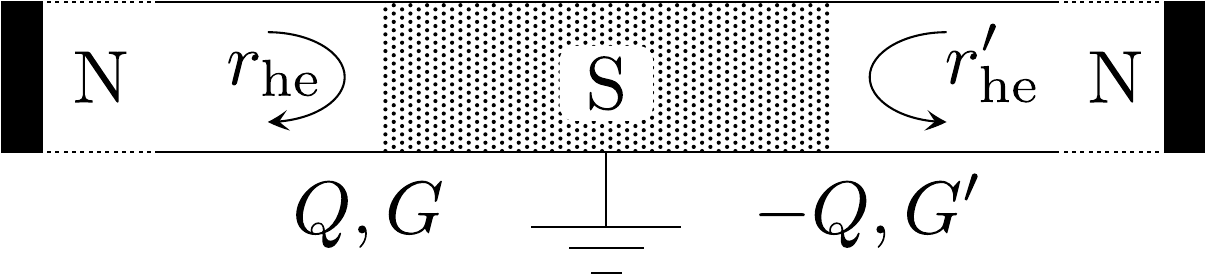}}
\caption{Superconducting wire (S) connected at both ends to a normal metal reservoir (N). The current $I$ flowing from the normal metal (at voltage $V$) into the grounded superconductor gives the conductance $G=I/V$ of the NS junction. The wire is assumed to be sufficiently long that there is negligible transmission from one end to the other. Chiral symmetry then produces a topologically protected quantum number $Q\in\mathbb{Z}$. Both $G=(2e^{2}/h){\rm Tr}\,r_{\rm he} r_{\rm he}^{\dagger}$ and $Q={\rm Tr}\,r_{\rm he}$ are determined by the Andreev reflection matrix $r_{\rm he}$ of the junction. While the NS junctions at the two ends of the wire can have independently varying conductances $G$ and $G'$, the topological quantum numbers are related by $Q'=-Q$. 
}
\label{fig:wire}
\end{figure}

In this paper we investigate the consequences of chiral symmetry for the electrical conductance of the superconducting nanowire, attached at the end to a normal metal contact. (See Fig.\ \ref{fig:wire}.) The conductance $G$ is determined by the matrix $r_{\rm he}$ of Andreev reflection amplitudes (from e to h, at the Fermi level), 
\begin{equation}
G=\frac{2e^{2}}{h}{\rm Tr}\,r_{\rm he}^{\vphantom{\dagger}} r_{\rm he}^{\dagger},\label{Grhedef}
\end{equation}
at low bias voltages and low temperatures and assuming that there is no transmission from one end of the wire to the other end. We will show that the topological quantum number $Q\in\mathbb{Z}$ in the presence of chiral symmetry is directly related to the Andreev reflection matrix,
\begin{equation}
Q={\rm Tr}\,r_{\rm he}.\label{Qrherelation}
\end{equation}
The intimate relation between transport and topology expressed by these two equations allows us to make specific predictions for the $Q$-dependence of $G$.

The outline of the paper is as follows. In the next section we derive Eq.\ \eqref{Qrherelation} from the general scattering formulation of one-dimensional topological invariants,\cite{Ful11} and obtain model-independent results for the relation between $G$ and $Q$. More specific results are obtained in Sec.\ \ref{sec:rmt} using random-matrix theory,\cite{Bee11} under the assumption of chaotic scattering at the NS interface. Then in Sec.\ \ref{sec:num} we numerically study a microscopic model of a superconducting nanowire,\cite{Lut10,Ore10} to test our analytical predictions in the presence of a realistic amount of chiral symmetry breaking. We conclude in Sec.\ \ref{conclude}.

\section{Relation between conductance and topological quantum number}
\label{scattering}

In a translationally invariant superconducting wire with chiral symmetry, the topological quantum number $Q$ can be extracted from the Bogoliubov-de Gennes Hamiltonian as a winding number in the one-dimensional Brillouin zone.\cite{Tew11} In order to make contact with transport measurements, we describe here an alternative scattering formulation for a finite disordered wire (adapted from Ref.\ \onlinecite{Ful11}), that expresses $Q$ as the trace of the Andreev reflection matrix from one of the ends of the wire. The electrical conductance $G$ can then be related to $Q$ by an inequality.

\subsection{Trace formula for the topological quantum number}
\label{symmQ}

The scattering problem is defined by connecting the $N$-mode superconducting wire (S) to a normal metal reservoir (N). The $2N\times 2N$ reflection matrix $r(E)$ relates the incident and reflected amplitudes of electron (e) and hole (h) excitations at energy $E$. It has a block structure of $N\times N$ submatrices,
\begin{equation}
  r=
  \begin{pmatrix}
    r_{\rm ee}  & r_{\rm eh} \\
    r_{\rm he} & r_{\rm hh}
  \end{pmatrix},\;\;\tau_{x}=\begin{pmatrix}
0&1\\
1&0
\end{pmatrix},
  \label{eq:r}
\end{equation}
where we have also introduced a Pauli matrix $\tau_{x}$ acting on the electron-hole degree of freedom.

We assume that both time-reversal symmetry and spin-rotation symmetry are broken, respectively, by a magnetic field and spin-orbit coupling. Electron-hole symmetry and chiral symmetry are expressed by
\begin{equation}
\tau_{x}r(-E)\tau_{x}=\left\{\begin{array}{cc}
r^{\ast}(E)&(\mbox{e-h symmetry}),\\
r^{\dagger}(E)&(\mbox{chiral symmetry}).
\end{array}\right.\label{rsymmetry}
\end{equation}
Taken together, the two symmetries imply that $r(E)=r^{T}(E)$ is a symmetric matrix. For spinless particles, this would be a time-reversal symmetry, but the true time-reversal symmetry also involves a spin-flip.

In what follows we consider the reflection matrix at the Fermi level ($E=0$). The symmetry relations \eqref{rsymmetry} then take the form
\begin{equation}
r_{\rm ee}=r_{\rm hh}^{\ast}=r_{\rm ee}^{T},\;\;r_{\rm he}=r_{\rm eh}^{\ast}=r_{\rm he}^{\dagger}.\label{rsymmetry0}
\end{equation}
These symmetries place the wire in universality class BDI, with topological quantum number determined\cite{Ful11} by the sign of the eigenvalues of the Hermitian matrix $\tau_{x}r$. This can be written as a trace if we assume that the wire is sufficiently long that we can neglect transmission of quasiparticles from one end to the other. The reflection matrix is then unitary, $rr^{\dagger}=1$. The matrix product $\tau_{x}r$ is both unitary and Hermitian, with eigenvalues $\pm 1$. The topological quantum number $Q\in\{-N,\ldots,-1,0,1,\ldots N\}$ is given by the trace
\begin{equation}
Q=\tfrac{1}{2}\,{\rm Tr}\,\tau_{x}r={\rm Tr}\,r_{\rm he}.\label{Qdef}
\end{equation}

All of this is for one end of the wire. The other end has its own reflection matrix $r'$, with topological quantum number $Q'=\frac{1}{2}{\rm Tr}\,\tau_{x}r'$. Unitarity with chiral symmetry relates $r$ and $r'$ via the transmission matrix $t$,
\begin{align}
&S=\begin{pmatrix}
r&t\\
t^{T}&r'
\end{pmatrix},\;\;(\tau_{x}S)^{2}=1,\nonumber\\
&\Rightarrow (\tau_{x}r)(\tau_{x}t)=-(\tau_{x}t)(\tau_{x}r').\label{rrprime}
\end{align}
If we allow for an infinitesimally small transmission for all modes through the wire, so that $t$ is invertible, this implies that ${\rm Tr}\,\tau_{x}r=-{\rm Tr}\,\tau_{x}r'$, hence $Q'=-Q$.

The sign of the topological quantum number at the two ends of the wire can be interchanged by a change of basis of the scattering matrix, $S\mapsto\tau_z S\tau_z$, so the sign of $Q$ by itself has no physical significance --- only relative signs matter.

\subsection{Conductance inequality}
\label{RAndreev}

In the most general case, the Andreev reflection eigenvalues $R_{n}\in[0,1]$ are defined as the eigenvalues of the Hermitian matrix product $r_{\rm he} r_{\rm he}^{\dagger}$. Because of chiral symmetry, the matrix $r_{\rm he}$ is itself Hermitian, with eigenvalues $\rho_{n}\in[-1,1]$ and $R_{n}=\rho_{n}^{2}$. These numbers determine the linear response conductance $G$ of the NS junction,
\begin{equation}
  G=G_0 \sum_{n=1}^N R_n, \;\;G_{0}=2e^{2}/h.
  \label{eq:G}
\end{equation} 
The factor of 2 in the definition of the conductance quantum $G_{0}$ is not due to spin (which is included in the sum over $n$), but accounts for the fact that charge is added to the superconductor as charge-$2e$ Cooper pairs.

The Andreev reflection eigenvalues $R_{n}$ different from $0,1$ are twofold degenerate (B\'{e}ri degeneracy).\cite{Ber09,Wim11} The eigenvalues $\rho_{n}$ are not degenerate, but another pairwise relation applies. Consider an eigenvalue $\rho$ of $r_{\rm he}$ with eigenvector $\psi$. It follows from the symmetry relations \eqref{rsymmetry0}, together with unitarity of $r$, that $r_{\rm he} r_{\rm hh}\psi^{\ast}=(r_{\rm eh} r_{\rm ee} \psi)^{\ast}=-(r_{\rm ee} r_{\rm he}\psi)^{\ast}=-\rho\, r_{\rm hh}\psi^{\ast}$. So $-\rho$ is  also an eigenvalue of $r_{\rm he}$, unless $r_{\rm hh}\psi^{\ast}=0$. This is not possible, again because of unitarity, if $|\rho|<1$. If also $\rho\neq 0$, the pair $\rho,-\rho$ is distinct.

So we see that the $\rho_{n}$'s different from $0,\pm 1$ come in pairs $\pm\rho$ of opposite sign. They cannot contribute to the topological quantum number $Q=\sum_{n}\rho_{n}$, only the $\rho_{n}$'s equal to $\pm 1$ can contribute (because $|Q|$ of them can come unpaired). Since each $|\rho_{n}|= 1$ contributes an amount $G_{0}$ to the conductance, we arrive at the lower bound
\begin{equation}
G/G_{0}\geq |Q|.\label{GQlower}
\end{equation}

The upper bound for $G/G_{0}$ is trivially $N$, the number of modes, but this can be sharpened if $N-|Q|$ is an odd integer. There must then be an unpaired $\rho_{n}=0$, leading to the upper bound
\begin{equation}
G/G_{0}\leq {\rm min}\bigl(N,N+(-1)^{N-|Q|}\bigr).\label{GQupper}
\end{equation}
For $N=1$ these inequalities imply $G/G_{0}=|Q|$, but for $N>1$ there is no one-to-one relationship between $G$ and $|Q|$.

Because the sign of $Q$ does not enter, the same inequalities constrain the conductances $G$ and $G'$ of the NS junctions at both ends of the wire (since $Q'=-Q$). Otherwise, the two conductances can vary independently.

Both inequalities \eqref{GQlower} and \eqref{GQupper}, derived here for symmetry class BDI with $|Q|=0,1,2,\ldots N$, apply as well to symmetry class D with $Q=0,1$ --- essentially because the B\'{e}ri degeneracy is operative there as well.\cite{Wim11}

\section{Conductance distribution for chaotic scattering}
\label{sec:rmt}

A statistical relation between conductance and topological quantum number can be obtained if we consider an ensemble of disordered wires and ask for the $Q$-dependence of the probability distribution $P(G)$. For chaotic scattering at the NS junction we can calculate the distribution from a circular ensemble of random-matrix theory. Such a calculation was performed in Ref.\ \onlinecite{Bee11} for a superconductor without chiral symmetry (symmetry class D). Here we follow that approach in the chiral orthogonal ensemble of symmetry class BDI.

The assumption of chaotic scattering requires a separation of time scales $\tau_{\rm dwell}\gg\tau_{\rm mixing}$, meaning that a quasiparticle dwells long enough at the NS interface for all available scattering channels to be fully mixed. Conceptually, this can be realized by confining the particles near the NS interface in a ballistic quantum dot.\cite{Bee11} In the next section we consider a microscopic model of a disordered NS interface with comparable dwell time and mixing time, but as we will see, the conductance distributions from the circular ensemble are still quite representative.

\subsection{Distribution of Andreev reflection eigenvalues}
\label{PofR}

We start from the polar decomposition of the reflection matrix in class BDI,
\begin{align}
  r =
  \begin{pmatrix}
    U & 0 \\ 0 & U^{\ast}
  \end{pmatrix}
  \begin{pmatrix}
    \Gamma &  \Lambda \\ \Lambda^{T} & \Gamma
  \end{pmatrix}
  \begin{pmatrix}
    U^{T} & 0 \\ 0 & U^{\dagger}
  \end{pmatrix}.\label{rpolar}
\end{align}
The matrix $U$ is an $N\times N$ unitary matrix and the $N\times N$ matrices $\Gamma,\Lambda$ are defined by
\begin{subequations}
\label{eq:polar}
\begin{align}
  \Gamma &= \bigoplus_{m=1}^{M}\begin{pmatrix}
\cos\alpha_{m}&0\\
0&\cos\alpha_{m}
\end{pmatrix}
\oplus \emptyset_{|Q|} \,\oplus\, \openone_\zeta  ,\label{eq:polara}\\
  \Lambda &= \pm\bigoplus_{m=1}^{M}\begin{pmatrix}
0&-i\sin\alpha_{m}\\
i\sin\alpha_{m}&0
\end{pmatrix} \oplus \openone_{|Q|}\,\oplus\, \emptyset_\zeta  .\label{eq:polarb}
\end{align}
\end{subequations}
The $\pm$ sign refers to the sign of $Q$. (For $Q=0$ the sign can be chosen arbitrarily.) The symbols $\openone_{n},\emptyset_{n}$ denote, respectively, an $n\times n$ unit matrix or null matrix. We have defined $\zeta =0$ if the difference $N-|Q|$ is even and $\zeta =1$ if $N-|Q|$ is odd. The $M=(N-|Q|-\zeta )/2$ angles $\alpha_m$ are in the interval $-\pi/2<\alpha_r\leq\pi/2$.

The Andreev reflection matrix $r_{\rm he}=(U\Lambda U^{\dagger})^{T}$ has eigenvalues $\rho_{n}=\sin\alpha_{n}$ ($n=1,2,\ldots M$), $\rho_{n}=-\sin\alpha_{n}$ ($n=M+1,M+2,\ldots 2M)$, $\rho_{n}=1$ ($n=2M+1,2M+2,\ldots 2M+|Q|$), and additionally $\rho_{N}=0$ if $N-|Q|$ is odd --- all of which is consistent with the general considerations of Sec.\ \ref{RAndreev}.

From the polar decomposition we obtain the invariant (Haar) measure $\mu(r)=r^{\dagger}dr$ that defines the uniform probability distribution in the circular ensemble, $P(r)d\mu(r)=d\mu(r)$. Upon integration over the independent degrees of freedom in the unitary matrix $U$ we obtain the distribution $P(\alpha_{1},\alpha_{2},\ldots\alpha_{M})$ of the angular variables. A change of variables then gives the distribution $P(R_{1},R_{2},\ldots R_{M})$ of the twofold degenerate Andreev reflection eigenvalues $R_{n}=\sin^{2}\alpha_{n}$. Details of this calculation are given in App.\ \ref{RMTcalculation}. The result is
\begin{equation}
	P(\{R_n\})\propto \prod_{m=1}^M {R_m}^{\zeta -1/2}(1-R_m)^{|Q|}\prod_{i<j=1}^M(R_i-R_j)^2.
	\label{eq:PofR}
\end{equation}

The $M$ twofold degenerate eigenvalues repel each other quadratically; furthermore, they are repelled with exponent $|Q|$ from the $|Q|$ eigenvalues pinned at unity. While the probability of finding a small reflection eigenvalue is enhanced for $N-|Q|$ even ($\zeta =0$), the eigenvalue $R_N=0$ pinned at zero for $N-|Q|$ odd ($\zeta =1$) produces a square root repulsion.

\subsection{Dependence of conductance distribution on the topological quantum number}
\label{PGresults}

Integration over the probability distribution \eqref{eq:PofR} of the Andreev reflection eigenvalues gives the distribution $P(g)$ of the dimensionless electrical conductance
\begin{align}
  g\equiv G/G_0=|Q|+2\sum_{m=1}^M R_m.
  \label{eq:cond}
\end{align}
The first term $|Q|$ is the quantized contribution from the topologically protected eigenvalues, and the factor of two in front of the sum accounts for the B\'{e}ri degeneracy of the $M$ eigenvalues without topological protection. 

\begin{figure}[tb] 
\centerline{\includegraphics[width=0.9\linewidth]{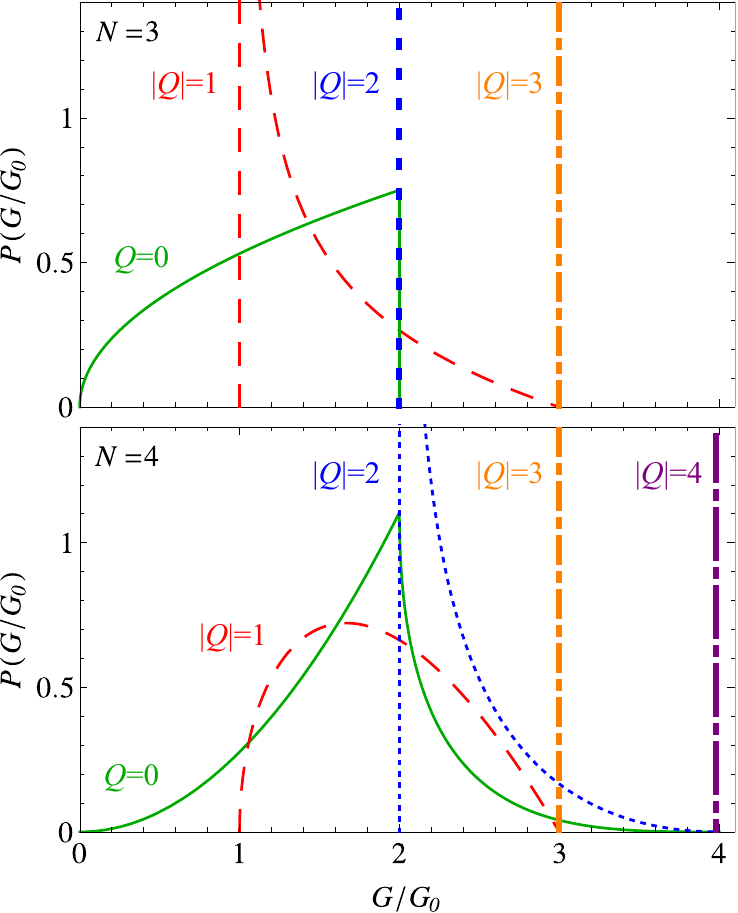}}
\caption{Probability distribution of the conductance for chaotic scattering in symmetry class BDI. The distributions are plotted from Eq.\ \eqref{Pgresults} for $N=3,4$ modes and different values of the topological quantum number $Q$. Thick vertical lines indicate a $\delta$-function distribution.
}
\label{fig:condDist}
\end{figure}

The conductance distribution is only nonzero in the interval 
\begin{equation}
|Q|\leq g\leq {\rm min}\bigl(N,N+(-1)^{N-|Q|}\bigr),\label{Qgineq}
\end{equation}
see Sec.\ \ref{RAndreev}. It is a trivial delta function, $P(g)=\delta(g-|Q|)$, when $|Q|=N,N-1$. Explicit results for small values of $N$ are
\begin{widetext}
\begin{subequations}
\label{Pgresults}
\begin{align}
&N=1:\; P(g)=\delta(g-|Q|),\\
&N=2:\; P(g)=\begin{cases}
\delta(g-|Q|)&\text{if}\;\;|Q|=1,2,\\
(8g)^{-1/2}&\text{if}\;\;|Q|=0,
\end{cases}\\
&N=3:\; P(g)=\begin{cases}
\delta(g-|Q|)&\text{if}\;\;|Q|=2,3,\\
\frac{3}{16}\sqrt{2}(3-g)(g-1)^{-1/2}\theta(g-1)&\text{if}\;\;|Q|=1,\\
\frac{3}{8}(2g)^{1/2}\theta(2-g)&\text{if}\;\;|Q|=0,\\
\end{cases}\\
&N=4:\;P(g)=\begin{cases}
\delta(g-|Q|)&\text{if}\;\;|Q|=3,4,\\
\frac{15}{128}\sqrt{2}(4-g)^2(g-2)^{-1/2}\theta(g-2)&\text{if}\;\;|Q|=2,\\
\frac{15}{32}\sqrt{2}(3-g)(g-1)^{1/2}\theta(g-1)\theta(3-g)&\text{if}\;\;|Q|=1,\\
\frac{45}{512}\pi g^2-\frac{45}{128} \bigl[ \sqrt{2}(4-g)\sqrt{g-2}+g^2\arctan\sqrt{\frac{1}{2}(g-2)} \bigr]\theta(g-2)&\text{if}\;\;|Q|=0.
\end{cases}\label{Pgresultsd}
\end{align}
\end{subequations}
\end{widetext}
The step function $\theta(x)$ (equal to 0 for $x<0$ and 1 for $x>0$) is used to indicate the nontrivial upper and lower bounds of the conductance. (The trivial bounds $0\leq g\leq N$ are not indicated explicitly.) The distributions for $N=3,4$ are plotted in Fig.\ \ref{fig:condDist}.

The first two moments of the conductance can be calculated in closed form for any value of $N,Q$, using known formulas for Selberg integrals.\cite{Sav06} (Alternatively, one can directly integrate over the BDI circular ensemble, see App.\ \ref{App_BDI_circular}.) We find
\begin{align}
  \langle G/G_0\rangle &= \frac{N(N-1)+Q^2}{2N-1} ,
  \label{eq:Gav} \\
  {\rm Var}\,(G/G_0) &= \frac{4(N^2-Q^2)(N^2-Q^2-2N+1)}{(2N-1)^2(2N+1)(2N-3)} ,
  \label{eq:varG}
  \end{align}
For $N\rightarrow\infty$ at fixed $|Q|$, this reduces to
\begin{align}
  \langle G/G_0\rangle &= \frac{N}{2}-\frac{1}{4}+\frac{Q^2-1/4}{2N} + \mathcal{O}(N^{-2}) ,
  \label{eq:Gavlarge} \\
  {\rm Var}\,(G/G_0) &= \frac{1}{4}-\frac{Q^2-1/4}{2N^2} + \mathcal{O}(N^{-3}).
  \label{eq:varGlarge}
\end{align}

The reduction of the average conductance below the classical value $NG_{0}/2=Ne^{2}/h$ is a weak localization effect, produced by the chiral symmetry in class BDI. (It is absent for the class-D circular ensemble.\cite{Alt97,Bee11}) The variance of the conductance in the large-$N$ limit, ${\rm Var}\,G\rightarrow (e^{2}/h)^{2}$, is twice as large as without chiral symmetry. 

A fundamental effect of chiral symmetry is that the $Q$-dependence of moments of the conductance is perturbative in $1/N$. In the class-D circular ensemble, in contrast, the $p$-th moment of the conductance is strictly independent of the topological quantum number for $N>p$, so topological signatures cannot be studied in perturbation theory.\cite{Bee11}

\section{Results for a microscopic model}
\label{sec:num}

We study a model Hamiltonian of a disordered two-dimensional semiconductor nanowire with induced superconductivity,\cite{Lut10,Ore10}
\begin{align}
  H ={}& \left(\frac{|\bm{p}|^2}{2m_{\rm eff}}+U(\bm{r})-\mu\right)\tau_z + V_Z\sigma_{x}\tau_z \nonumber\\
  &+ \frac{\alpha_{\rm so}}{\hbar}(p_x\sigma_y\tau_z - p_y\sigma_x)  +\Delta_0\sigma_y\tau_y.
  \label{eq:Hnanowire}
\end{align}
This Bogoliubov-de Gennes Hamiltonian contains the single-particle kinetic energy $(p_x^2+p_y^2)/2m_{\rm eff}$, electrostatic disorder potential $U(x,y)$, chemical potential $\mu$, Zeeman energy $V_{\rm Z}$, Rashba spin-orbit coupling constant $\alpha_{\rm so}$, and \textit{s}-wave pairing potential $\Delta_0$. The Pauli matrices $\sigma_{i}$, $\tau_{i}$ act on the spin and electron-hole degree of freedom, respectively. The two-dimensional wire has width $W$ in the $y$-direction and extends along the $x$-direction (parallel to the Zeeman field). We define the spin-orbit coupling length $l_{\rm so} = \hbar^{2}(m_{\rm eff}\alpha_{\rm so})^{-1}$ and confinement energy $E_W=\hbar^{2}(2m_{\rm eff}W^{2})^{-1}$.

\subsection{Mechanisms for chiral symmetry breaking}
\label{chiralmechanisms}

Electron-hole symmetry and chiral symmetry,
\begin{equation}
\tau_{x}H\tau_{x}=\left\{\begin{array}{cc}
-H^{\ast}&(\mbox{e-h symmetry}),\\
-H&(\mbox{chiral symmetry}),
\end{array}\right. \label{Hsymmetry}
\end{equation}
together require that $H$ is real. While the electron-hole symmetry is an exact symmetry of the Hamiltonian \eqref{eq:Hnanowire}, the chiral symmetry is broken by the spin-orbit term $p_y\sigma_x$ associated with transverse motion.\cite{Tew11} 
%If $U=\mu=0$ one can still recover the chiral symmetry in the presence of the transverse spin-orbit term,\cite{Che10} but that is not a relevant limit here.

To quantify the stability of multiple zero-energy states, we follow Ref.~\onlinecite{Scheid09} and make a unitary transformation $H\mapsto {\cal U}^{\dagger}H{\cal U}\equiv H'$ with
${\cal U} =  \exp(i\sigma_x\tau_z\,y/l_\text{so})$. The transformed Hamiltonian,
\begin{equation}
\begin{split}
H' ={}& \left(
\frac{|\bm{p}|^2}{2m_\text{eff}} +U -\frac{\alpha_{\rm so}}{2l_{\rm so}}
-\mu\right) \tau_z + V_Z \sigma_x \tau_z+ \Delta_0 \sigma_y \tau_y \\
& +  \frac{\alpha_\text{so}}{\hbar} p_x[\cos(2y/l_\text{so}) \sigma_y \tau_z
+ \sin(2y/l_\text{so}) \sigma_z]
,
\end{split}
\end{equation}
no longer contains $p_{y}$ and breaks chiral symmetry through the final term $\propto p_x\sigma_z$. For $W\ll l_{\rm so}$ this term produces a splitting $\delta E$ of pairs of zero-energy states of order $(W/l_\text{so}) E_\text{gap}$, with $E_\text{gap} \propto
\alpha_\text{so}$ the induced superconducting gap. This simple estimate is an upper bound on the splitting, even smaller splittings have been found in Refs.\ \onlinecite{Kells12,Potter12, Rieder12}. We typically find in our numerical simulations that $\delta E\lesssim 0.05\,E_{\rm gap}$ for $W\lesssim l_{\rm so}$.
%In the limit of wider wires, $W/l_\text{so} \gtrsim
%1$, the splitting is typically a sizable fraction of the gap (from
%numerical simulations we find $\sim 0.05-0.5 E_\text{gap}$ for
%$1<W/l_\text{so}<3$).

There are other methods to break chiral symmetry. An externally controllable method is to tilt the magnetic field so that it acquires a nonzero component in the $y$-direction, in the plane of the substrate but perpendicular to the axis of the nanowire.\cite{Tew12} The orbital effect of a magnetic field (Lorentz force) also breaks chiral symmetry, but this is expected to be small compared to the Zeeman effect on the spin. Subband-mixing by a disorder potential or a position-dependent pairing term preserve chiral symmetry.
This leaves spin-orbit
coupling of transverse momentum as the most significant intrinsic mechanism for chiral symmetry breaking
and we will focus on it in the simulations. We find a transition from symmetry class D ($Q\in\mathbb{Z}_2$) to class BDI ($Q\in\mathbb{Z}$) if $W$ drops below $l_{\rm so}$.

All these considerations apply to noninteracting quasiparticles. Interactions have the effect of restricting $Q$ to $\mathbb{Z}_{8}$, so chiral symmetry can stabilize at most $8$ zero-modes at each end of the wire.\cite{Fid11,Lut11} For $N\leq 8$ we expect the universal class BDI results (in particular the conductance quantization) to be unaffected by interactions.

\subsection{Class BDI phase diagram}
\label{phasediagram}

\begin{figure}[tb] 
\centerline{\includegraphics[width=1\linewidth]{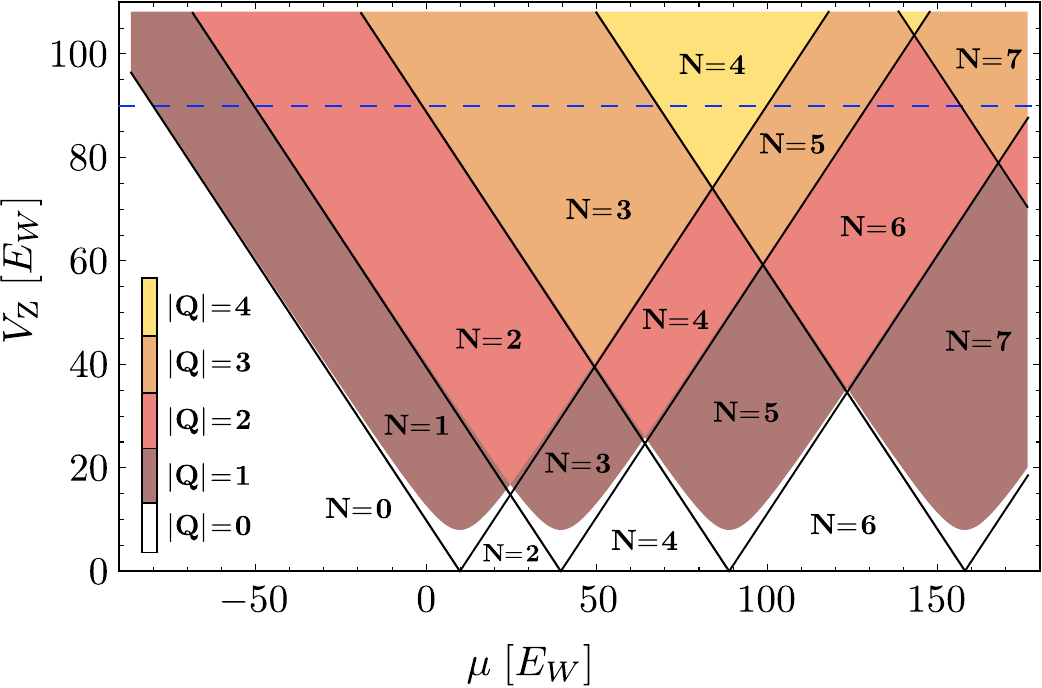}}
	\caption{Topological phase diagram of the Hamiltonian \eqref{eq:Hnanowire} without disorder ($U\equiv 0$) and without any chiral symmetry breaking ($\alpha p_y\sigma_x\equiv 0$, symmetry class BDI). The colored regions give the value of the topological quantum number $Q$ in the superconducting state ($\Delta_{0}=8E_{W}$), while the black lines separate regions with different number of propagating modes $N$ in the normal state ($\Delta_{0}=0$). The topological phase boundaries are independent of $E_{\rm so}$.
The blue line is referred to in Fig.\ \ref{fig:steps}.
}
	\label{fig:phases}
\end{figure}

For an infinite clean wire with exact chiral symmetry, Fig.\ \ref{fig:phases} shows the phases with different topological quantum number $Q\in\mathbb{Z}$ as a function of Zeeman energy and chemical potential. (A similar phase diagram is given in Ref.\ \onlinecite{Tew12}.) The phase boundaries are determined from the Hamiltonian \eqref{eq:Hnanowire} by setting $\alpha p_y\sigma_x\equiv 0$, $U\equiv 0$, and demanding that the excitation gap vanishes. This happens at
\begin{align}
&p_x=0,\;\;p_y=p_n=n\pi\hbar/W,\;\;n=1,2,\ldots N,\nonumber\\
&V_{\rm Z}^{2}=\Delta_{0}^{2}+(\mu-p_n^{2}/2m_{\rm eff})^{2},\label{gapclosing}
\end{align}
with $N$ the number of propagating modes in the normal state ($\Delta_{0}=0$).

If one follows the sequence of $Q,N$ values with increasing $\mu$ at constant $V_{\rm Z}$, one sees that $|Q|$ remains equal to $N\geq 1$ for a range of chemical potentials $(\mu-\pi^{2} E_W)^{2}\lesssim V_{\rm Z}^{2}$. For example, the sequence along the dashed blue line is $(|Q|,N)=(1,1),(2,2),(3,3),(4,4),\ldots$. In view of the inequality \eqref{GQlower}, this implies a sequence of $2e^2/h$ conductance steps. 
The first quantized conductance plateau emerges when the Zeeman energy exceeds the superconducting gap ($V_Z>\Delta_0$). Additional plateaus form at fields, for which the Zeeman energy becomes larger than the subband splitting. More specifically, the $n$-th conductance plateau appears for $V_Z^2 = \Delta_0^2 + E_W^2\pi^4(n^2-1)^2/4$ ($n=1,2,3,\dots$).

\subsection{Conductance quantization}
\label{Gquant}

\begin{figure}[tb] 
\centerline{\includegraphics[width=1\linewidth]{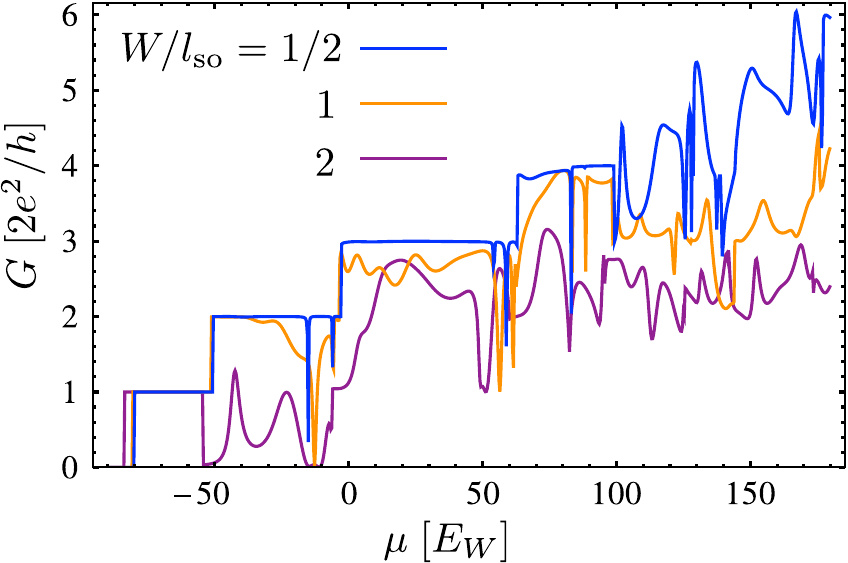}}
	\caption{Conductance of a disordered NS junction, calculated numerically from the model Hamiltonian \eqref{eq:Hnanowire}. The chemical potential $\mu$ is increased at constant $V_{\rm Z}=90\,E_{W}$, $\Delta_{0}=8E_{W}$ (blue dashed line in the BDI phase diagram of Fig.~\ref{fig:phases}), for three different values of the spin-orbit coupling length $l_{\rm so}$. Each curve is for a single disorder realization (of strength $U_0=180\,E_{W}$). The conductance quantization at $2,3,4\times 2e^2/h$ is lost by chiral symmetry breaking as $W$ becomes larger than $l_{\rm so}$.
}
\label{fig:steps}
\end{figure}

To demonstrate the conductance quantization we attach a clean normal-metal lead at $x=0$ to the disordered superconducting wire. For $x<0$ we thus set $\Delta_{0}=0$ and $U=0$. The Andreev reflection matrix is calculated numerically by discretizing the Hamiltonian on a square lattice (lattice constant $a=W/20$). Disorder is realized by an electrostatic potential $U(x,y)$ that varies randomly from site to site for $x>0$, distributed uniformly in the interval $(-U_0,U_0)$.

The results in Fig.\ \ref{fig:steps} clearly show the expected behavior: For $W= l_{\rm so}/2$ the conductance increases in a sequence of quantized steps, insensitive to disorder, as long as $|Q|\in\{N,N-1\}$. The quantization at $|Q|\geq 2$ is lost for $W=2l_{\rm so}$ because of chiral symmetry breaking. The very first step $G=2e^{2}/h$ is common to both symmetry classes D and BDI, so it persists.

\subsection{Conductance distribution}
\label{conddistr}

\begin{figure}[tb] 
\centerline{\includegraphics[width=0.9\linewidth]{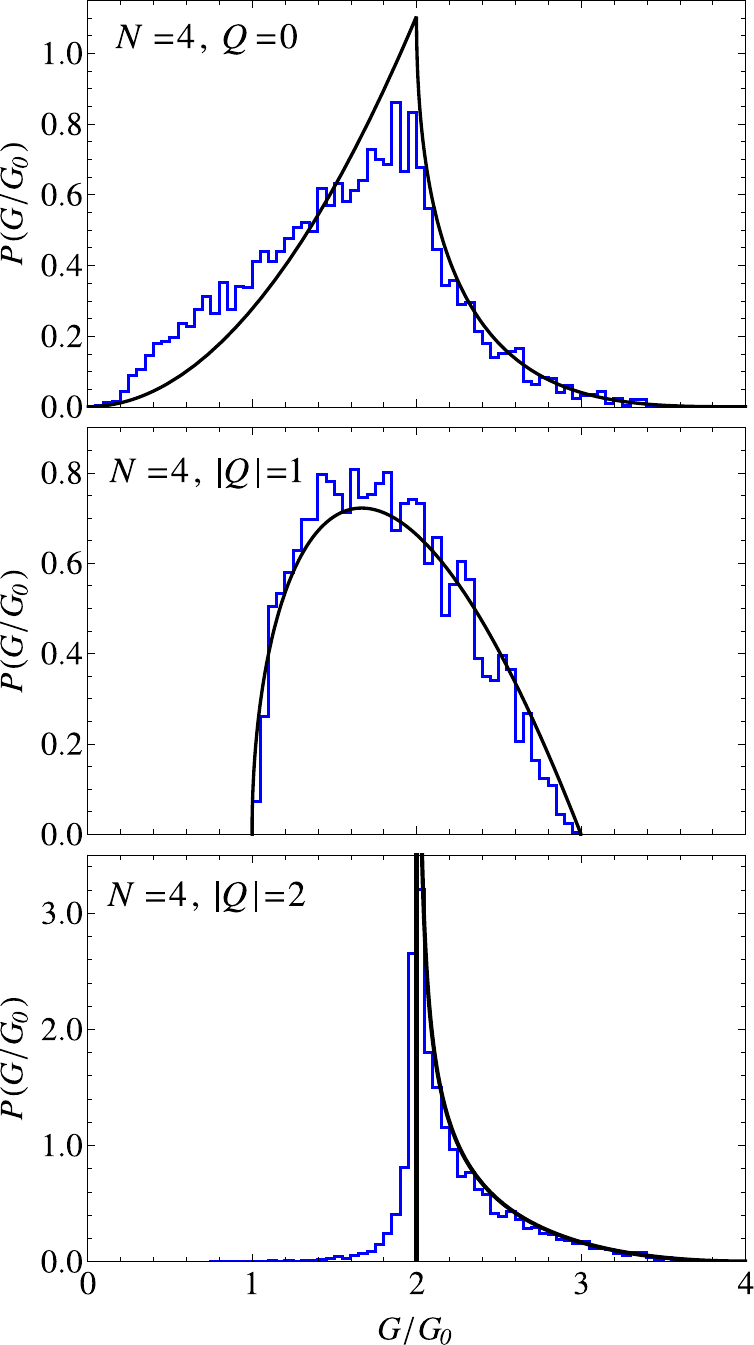}}
\caption{Blue histograms: probability distribution of the conductance of the NS junction, calculated from the model Hamiltonian \eqref{eq:Hnanowire} in an ensemble of disorder realizations. Each panel has the same number of modes $N=4$ in the normal region and a different topological quantum number $|Q|=0,1,2$ in the superconductor. The black curves are the corresponding distributions in the class BDI circular ensemble, given by Eq.\ \eqref{Pgresultsd}. Each panel has the same value of $l_{\rm so}=2W$ and $\Delta_{0}=8E_W$. The other energy scales (in units of $E_W$) are as follows:
$\bm{Q=0}$: $\mu_N=\mu_S=64$, $V_{\rm Z}=14$, $U_0=180$; $\bm{|Q|=1}$: $\mu_N=64$, $\mu_S=88$, $V_{\rm Z}=14$, $U_0=180$; $\bm{|Q|=2}$: $\mu_N=\mu_S=64$, $V_{\rm Z}=34$, $U_0=140$.
}
\label{fig:n4_stat}
\end{figure}

For $|Q|\leq N-2$ there is no conductance quantization, but we can still search for the $Q$-dependence in the statistical distribution of the conductance. In Fig.\ \ref{fig:n4_stat} we show the distribution function for $N=4$, $|Q|=0,1,2$, calculated by averaging the results of the numerical simulation over disorder realizations. The parameters used are listed in the caption. The values of the Fermi energy ($\mu_{N}$ in the normal region and $\mu_{S}$ in the superconducting region) were chosen in order to be far from boundaries where $Q$ or $N$ changes. 

We found that the conductance distributions depend sensitively on the disorder strength, demonstrating that the scattering at the NS interface is diffusive rather than chaotic. This is as expected, since chaotic scattering requires a confined geometry (for example, a quantum dot), to fully mix the scattering channels. Still, by adjusting the disorder strength $U_{0}$ a quite good agreement could be obtained with the distribution from the class BDI circular ensemble calculated in Sec.\ \ref{PGresults}. Since this is a single fit parameter for an entire distribution function, we find the agreement with the circular ensemble quite satisfactory.

\section{Conclusion}
\label{conclude}

In conclusion, we have developed a scattering theory for superconducting nanowires with chiral symmetry (symmetry class BDI), relating the electrical conductance $G$ to the topological quantum number $Q\in\mathbb{Z}$. In a closed system $|Q|$ counts the number of Majorana zero-modes at the end of the wire, but in our open system these end states have broadened into a continuum with other nontopological states. Still, the value of $|Q|$ manifests itself in the conductance as a quantization $G=|Q|\times 2e^{2}/h$ over a range of chemical potentials (see Fig.\ \ref{fig:steps}). 

More generally, even when $G$ is not quantized, the conductance distribution is sensitive to the value of $|Q|$, as we calculated in the circular ensemble of random-matrix theory (see Fig.\ \ref{fig:n4_stat}). Comparison with Ref.\ \onlinecite{Bee11}, where the conductance distribution was calculated in the absence of chiral symmetry (symmetry class D with $Q\in\mathbb{Z}_{2}$), shows that chiral symmetry manifests itself even when $|Q|\leq 1$ --- so even if there is not more than a single Majorana zero-mode.

The chiral symmetry is an approximate symmetry (unlike the fundamental electron-hole symmetry), requiring in particular a wire width $W$ below the spin-orbit coupling length $l_{\rm so}$. Our model calculations in Fig.\ \ref{fig:steps} show that chiral symmetry is lost for $W\gtrsim 2l_{\rm so}$ and well preserved for $W\lesssim l_{\rm so}/2$. Existing experiments\cite{Mou12,Den12,Das12} on InAs and InSb nanowires typically have $l_{\rm so}\simeq 200\,{\rm nm}$ and $W\simeq 100\,{\rm nm}$. These are therefore in the chiral regime and can support more than a single zero-mode at each end, once the Zeeman energy becomes comparable to the subband spacing.

\acknowledgments

The numerical simulations of the nanowire were performed with the {\sc kwant} software package, developed by A. R. Akhmerov, C. W. Groth, X. Waintal, and M. Wimmer. We gratefully acknowledge discussion with I.~Adagideli.
Our research was supported by the Dutch Science Foundation NWO/FOM, by an ERC Advanced Investigator Grant, and by the EU network NanoCTM.

\appendix

\section{Calculation of the Andreev reflection eigenvalue distribution in the BDI circular ensemble}
\label{RMTcalculation}

In this Appendix we derive the probability distribution $P(\{R_m\})$ of the B\'{e}ri degenerate Andreev reflection eigenvalues $R_1,\dots,R_M$ in the circular ensemble of symmetry class BDI (circular chiral orthogonal ensemble). The calculation follows the standard procedure of random-matrix theory,\cite{For10} and is technically similar to the calculation for symmetry class D (circular real ensemble) presented in Ref.\ \onlinecite{Bee11}.

The probability distribution $P(\{R_m\})$ is determined by the invariant (Haar) measure $d\mu(r)=r^\dagger dr=\delta r$, which for a given topological quantum number $Q$ characterizes the uniform distribution of scattering matrices in the circular ensemble subject to the symmetry constraints of Eq.~(\ref{rsymmetry0}). 
Since any scattering matrix in the ensemble can be decomposed according to Eq.~(\ref{rpolar}), i.e.\ parameterized in terms of the angles $\alpha_m$, we can transform the invariant measure into $d\mu(r)=J\prod_i dp_i\prod_m d\alpha_m$. 
The $p_i$'s denote the degrees of freedom of the matrix of eigenvectors $U$ and $J$ is the Jacobian of the transformation. 
From this expression the distribution of the angles $\alpha_m$ follows via integration over the $p_i$'s. Up to a normalization constant we have
\begin{equation}
  P(\{\alpha_m\})\propto\int J\prod_i dp_i \,.
  \label{eq:palphardev}
\end{equation}

The polar decomposition in Eq.~(\ref{rpolar}) is not unique. As in Ref.\ \onlinecite{Bee11} the redundant degrees of freedom can be removed by restricting the independent parameters $p_i$ in the matrix of eigenvectors $U$.
The total number of degrees of freedom furthermore depends on $N$ as well as on $Q$.
This is best seen if one considers the reflection matrix $\tilde r$ in a basis where it is a real orthogonal and symmetric matrix, of the form
\begin{align}
  \tilde r = O
  \begin{pmatrix}
    \openone_{N+Q} & 0 \\ 0 & -\openone_{N-Q} 
  \end{pmatrix}
  O^T ,
  \label{eq:OrtildeO}
\end{align}
with $O$ a $2N\times 2$N real orthogonal matrix.
In this basis the topological quantum number is given by $Q=\frac{1}{2}\,{\rm Tr}\,\tilde r$.
The upper-left and lower-right blocks do not change under an additional orthogonal transformation $O'_{N+Q}\oplus O''_{N-Q}$. 
Group division readily gives the total number of degrees of freedom: $\operatorname{dim}O(2N)-\operatorname{dim}O(N+Q)-\operatorname{dim}O(N-Q)=N^2-Q^2$. 
Since there are $M$ angular parameters $\alpha_{m}$, there must be $N^2-Q^2-M$ independent degrees of freedom $p_i$ in the matrix of eigenvectors $U$.

In order to obtain the probability distribution from Eq.~(\ref{eq:palphardev}) we need the Jacobian $J$.
It can be determined from the metric tensor $g_{\mu\nu}$, which can be extracted from the trace ${\rm Tr}\,\delta r\delta r^\dagger$, when it is expressed in terms of the infinitesimals $d\alpha_m$ and $dp_i$ (collectively denoted as $dx_\mu$):
\begin{equation}
  {\rm Tr}\,\delta r\delta r^\dagger=\sum_{\mu,\,\nu}g_{\mu\nu}dx_\mu dx_\nu \,.
  \label{eq:traceJ}
\end{equation}

In view of the polar decomposition (\ref{rpolar}) one has
\begin{equation}
  W^\dagger dr W^* =\delta WL+dL+L\delta W^T,
  \label{eq:WdrW}
\end{equation}
where we abbreviated
\begin{equation}
  W=
  \begin{pmatrix}
    U & 0 \\ 0 & U^*
  \end{pmatrix},\;\;
  L=
  \begin{pmatrix}
    \Gamma & \Lambda \\ \Lambda^T & \Gamma
  \end{pmatrix} \,.
  \label{eq:WandL}
\end{equation}
Unitarity ensures $0=d(U^\dagger U)=dU^\dagger U+U^\dagger dU\Rightarrow\delta U^\dagger=-\delta U$.
Substitution of Eqs.~(\ref{eq:WdrW}) and (\ref{eq:WandL}) into ${\rm Tr}\,\delta r\delta r^\dagger={\rm Tr}\, drdr^\dagger={\rm Tr}\,(W^\dagger dr W^*W^Tdr^\dagger W)$ gives
\begin{align}
  {\rm Tr}\,\delta r\delta r^\dagger = {\rm Tr}\,(dLdL^\dagger-2L\delta W^TL^\dagger\delta W-2\delta W^2)\,.
\end{align}
From the block structure of $W$ and $L$ we find ${\rm Tr}\,\delta W^2=2\,{\rm Tr}\,\delta U^2$ and
\begin{equation}
  {\rm Tr}\,(L\delta W^T L^\dagger\delta W) = 2\,{\rm Tr}\,(\Gamma\delta U\Gamma\delta U^T-\Lambda\delta U\Lambda\delta U),
\end{equation}
where we have used $\Gamma^T=\Gamma^*=\Gamma$ and $\Lambda^\dagger=\Lambda$.

It is convenient to express ${\rm Tr}\,\delta r\delta r^\dagger$ in terms of the form ${\rm Tr}\, AA^\dagger=\sum_{ij}|A_{ij}|^2$. 
Using $\Gamma\Gamma +\Lambda\Lambda=\openone_N$ we find
\begin{align}
  {\rm Tr}\,\delta r\delta r^\dagger &= {\rm Tr}\, dLdL^\dagger 
  + 2{\rm Tr}\,(\Gamma\delta U^T+\delta U\Gamma)(\Gamma\delta U^T+\delta U\Gamma)^\dagger \nonumber \\
  &\quad+ 2{\rm Tr}\,(\Lambda\delta U-\delta U\Lambda)(\Lambda\delta U-\delta U\Lambda)^\dagger \nonumber \\
  &\equiv T_1 + T_2 + T_3.
  \label{eq:T1to3}
\end{align}
The first trace simply evaluates to
\begin{equation}
  T_1=2\sum_{ij=1}^N\bigl(|d\Gamma_{ij}|^2+|d\Gamma_{ij}|^2\bigr) =4\sum_{m=1}^M(d\alpha_m)^2 \,.
  \label{eq:T1}
\end{equation}

The remaining two traces $T_2$ and $T_3$ need to be calculated using the block structure of $\Gamma$ and $\Lambda$ in Eq.~(\ref{eq:polar}).
We work out the calculation for $Q=0$, $N$ even ($\Rightarrow\zeta =0$). 
The two matrices $\Lambda$ and $\Gamma$ are in this case fully described by $M=N/2$ blocks of $2\times 2$ matrices.
The two traces evaluate to
\begin{widetext}
\begin{align}
	\tfrac{1}{4}T_2 = &\sum_{k=1}^M 2\cos^2\alpha_k\left\{ |\delta U_{2k,2k}|^2+|\delta U_{2k-1,2k-1}|^2+2\left[{\rm Im}\,(\delta U_{2k-1,2k})\right]^2\right\} \nonumber\\
  &+\sum_{k<{l}=1}^M\left\{ (\cos^2\alpha_k+\cos^2\alpha_{l})(|\delta U_{2k,2{l}}|^2+|\delta U_{2k,2{l}-1}|^2+|\delta U_{2k-1,2{l}}|^2+|\delta U_{2k-1,2{l}-1}|^2)\right. \nonumber\\
  &\left.\qquad\quad\:-2\cos\alpha_l\cos\alpha_{k}\,{\rm Re}\,(\delta U_{2l,2{k}}^2+\delta U_{2l,2{k}-1}^2+\delta U_{2l-1,2{k}}^2+\delta U_{2l-1,2{k}-1}^2)\right\},
  \label{eq:t2}
\end{align}
\begin{align}
	\tfrac{1}{4}T_3 = &\sum_{k=1}^M \sin^2\alpha_k\left\{|\delta U_{2k,2k}-\delta U_{2k-1,2k-1}|^2 
  + 4\left[{\rm Im}\,(\delta U_{2k-1,2k})\right]^2\right\} \nonumber\\
  &+ \sum_{k<{l}=1}^M \left\{(\sin^2\alpha_k+\sin^2\alpha_{l})(|\delta U_{2k,2{l}}|^2+|\delta U_{2k,2{l}-1}|^2+|\delta U_{2k-1,2{l}}|^2+|\delta U_{2k-1,2{l}-1}|^2)\right. \nonumber\\
  &\left.\qquad\quad\:+\,4\sin\alpha_k\sin\alpha_{l}\,{\rm Re}\,(\delta U_{2k,2{l}-1}\delta U_{2k-1,2{l}}^*-\delta U_{2k,2{l}}\delta U_{2k-1,2{l}-1}^*)\right\}.
  \label{eq:t3}
\end{align}

Like $\Gamma$ and $\Lambda$ the elements of the matrix $\delta U$ can be grouped into separate $2\times 2$ blocks, denoted by the block indices $k, l=1,\dots,M$.
We first consider the block-off-diagonal part for which we can choose as independent parameters
\begin{align}
	\delta U_{2k,2l},\; \delta U_{2k,2l-1},\; \delta U_{2k-1,2l},\; \delta U_{2k-1,2l-1}, \nonumber
\end{align}
with $1\leq k < l\leq M$. 
The real and imaginary parts, denoted by $\delta U^\text{R}, \delta U^\text{I}$, produce a total of $4M(M-1)$ independent parameters.
Note that $\delta U^\dagger = -\delta U$ immediately implies $\delta U_{2k,2l}^\text{R}=-\delta U_{2l,2k}^\text{R}$, $\delta U_{2k,2l}^\text{I}=\delta U_{2l,2k}^\text{I}$, and so on.
For given values of $k$ and $l$ the contribution to ${\rm Tr}\,\delta r\delta r^\dagger$ has the form
\begin{align}
	&a\!\left[(\delta U_{2k,2l}^\text{R})^2\!\! +\! (\delta U_{2k,2l-1}^\text{R})^2\!\! +\! (\delta U_{2k-1,2l}^\text{R})^2\!\! + \!(\delta U_{2k-1,2l-1}^\text{R})^2 \right] 
	+ b\!\left[(\delta U_{2k,2l}^\text{I})^2\!\! +\! (\delta U_{2k,2l-1}^\text{I})^2\!\! +\! (\delta U_{2k-1,2l}^\text{I})^2\!\! +\! (\delta U_{2k-1,2l-1}^\text{I})^2 \right] \nonumber\\
	&\!+\! 2c\big[\delta U_{2k,2l-1}^\text{R}\delta U_{2k-1,2l}^\text{R} + \delta U_{2k,2l-1}^\text{I}\delta U_{2k-1,2l}^\text{I}  
	- \delta U_{2k,2l}^\text{R}\delta U_{2k-1,2l-1}^\text{R} - \delta U_{2k,2l}^\text{I}\delta U_{2k-1,2l-1}^\text{I}\big]\;,
\nonumber
\end{align}
where we abbreviated $a=2(1-\cos\alpha_k\cos\alpha_l)$, $b=2(1+\cos\alpha_k\cos\alpha_l)$, and $c=2\sin\alpha_k\sin\alpha_l$.
\end{widetext}

The contribution to the metric tensor is a block matrix
\begin{align}
	\begin{pmatrix}
		a & -c & 0 & 0 \\
		-c & a & 0 & 0 \\
		0 & 0 & a & c \\
		0 & 0 & c & a
	\end{pmatrix}
	\oplus
	\begin{pmatrix}
		b & -c & 0 & 0 \\
		-c & b & 0 & 0 \\
		0 & 0 & b & c \\
		0 & 0 & c & b
	\end{pmatrix} \;,
\nonumber
\end{align}
where the first and the second block correspond to the real and imaginary parts, respectively.
The determinant of this block matrix is $256\,(\sin^2\alpha_k-\sin^2\alpha_l)^4$. This gives us the contribution to the Jacobian from the off-diagonal matrix elements
\begin{align}
	J_\text{off-diagonal}=\prod_{k<l=1}^M (\sin^2\alpha_k-\sin^2\alpha_l)^2.
	\label{eq:Joff}
\end{align}

Next  we consider the diagonal $2\times 2$ blocks. Anti-Hermiticity of $\delta U$ implies $\delta U^\text{R}_{ii}=0$ ($i=1,\dots,N$). We choose the $3M$ independent parameters
\begin{align}
	\delta U_{2k,2k}^\text{I},\quad \delta U_{2k-1,2k-1}^\text{I},\quad \delta U_{2k-1,2k}^\text{I} \;.
\nonumber
\end{align}
The contribution to ${\rm Tr}\,\delta r\delta r^\dagger$ for a given value $k$ has the form
\begin{align}
	& v\left[(\delta U_{2k,2k}^\text{I})^2 + (\delta U_{2k-1,2k-1}^\text{I})^2\right] \nonumber\\
	& -2w\delta U_{2k,2k}^\text{I}\delta U_{2k-1,2k-1}^\text{I} + 4(\delta U_{2k-1,2k}^\text{I})^2 \;,
\nonumber
\end{align}
where $v=1+\cos^2\alpha_k$ and $w=\sin^2\alpha_k$. Note that $\delta U_{2k-1,2k}^\text{I}$ is fully decoupled. The contribution to the metric tensor is
\begin{align} 
	\begin{pmatrix}
		v & -w \\ -w & v
	\end{pmatrix}
\nonumber
\end{align}
with a determinant of $4\cos^2\alpha_k$. This leads to a contribution to the Jacobian from the diagonal matrix elements
\begin{align}
  J_\text{diagonal} = \prod_{k=1}^M (1-\sin^2\alpha_k)^{1/2} \;.
	\label{eq:Jdiagonal}
\end{align}

The total number of independent parameters that we have accounted for is $4M^2=N^2$ (including the $M$ angular parameters $\alpha_m$). This is exactly the number we expect for $N$ even and $Q=0$. 
Collecting all the terms that contribute to the Jacobian in Eq.~(\ref{eq:palphardev}), we obtain the probability distribution
\begin{align}
	P({\alpha_k}) \propto \prod_{k=1}^M(1-\sin^2\alpha_k)^{1/2} \prod_{k<l=1}^M(\sin^2\alpha_k-\sin^2\alpha_l)^2.
	\label{eq:palpha}
\end{align}
Integration over the $N^2-|Q|^2-M$ ancillary degrees of freedom of the matrix of eigenvectors $U$ only gives rise to an overall constant. A transformation of variables from $\alpha_m$ to $R_m=\sin^2\alpha_m$ gives the distribution \eqref{eq:PofR} of the twofold degenerate Andreev reflection values in the case $Q=0$, $N$ even ($\zeta=0$).
The cases $Q\neq 0$ and/or $N$ odd are worked out similarly. 

\section{Average conductance in the BDI circular ensemble}
\label{App_BDI_circular}

In the circular ensemble of Sec.\ \ref{PGresults} the $2N\times 2N$ reflection matrix $r$ is uniformly distributed in the unitary group, subject to the restrictions of electron-hole symmetry and chiral symmetry. The average conductance can be calculated directly by integration over the unitary group. We give this calculation here, as a check on the result \eqref{eq:Gav} derived by going through the distribution of Andreev reflection eigenvalues.

Unitarity ($rr^{\dagger}=1$) implies that the expression \eqref{Grhedef} for the conductance can be written equivalently as
\begin{align}
G&=\tfrac{1}{4}G_{0}\,{\rm Tr}\,\bigl(1+r_{\rm he}^{\vphantom{\dagger}} r_{\rm he}^{\dagger}+r_{\rm eh}^{\vphantom{\dagger}} r_{\rm eh}^{\dagger}-r_{\rm ee}^{\vphantom{\dagger}} r_{\rm ee}^{\dagger}-r_{\rm hh}^{\vphantom{\dagger}} r_{\rm hh}^{\dagger}\bigr)\nonumber\\
&=\tfrac{1}{4}G_{0}\,{\rm Tr}\,\bigl(1-\tau_{z}r\tau_{z}r^{\dagger}\bigr).\label{Grdef}
\end{align}
Electron-hole symmetry ($r=\tau_{x}r^{\ast}\tau_{x}$) and chiral symmetry ($r=r^{T}$) constrain $r$ to the form
\begin{equation}
r=-ie^{i\tau_{x}\pi/4}O{\cal D}_{Q}O^{T}e^{i\tau_{x}\pi/4}.\label{rOdef}
\end{equation}
The matrix $O$ is real orthogonal ($OO^{T}=1$). The diagonal matrix ${\cal D}_{Q}$ has entries $\pm 1$ on the diagonal with ${\rm Tr}\,{\cal D}_{Q}=2Q$, consistent with Eq.\ \eqref{Qdef}. Substitution into Eq.\ \eqref{Grdef} gives
\begin{equation}
G=\tfrac{1}{4}G_{0}\,{\rm Tr}\,\bigl(1+\tau_{y}O{\cal D}_{Q}O^{T}\tau_{y}O{\cal D}_{Q}O^{T}\bigr).\label{Grdef2}
\end{equation}

In the circular ensemble the matrix $O$ is uniformly distributed with respect to the Haar measure for $2N\times 2N$ orthogonal matrices. The average of a product of four orthogonal matrices equals\cite{Col06}
\begin{widetext}
\begin{align}
\langle O_{\alpha a}O_{\beta b}O_{\gamma c}O_{\delta d}\rangle={}&\frac{2N+1}{2N(2N-1)(2N+2)}\bigl(\delta_{\alpha\beta}\delta_{ab}\delta_{\gamma\delta}\delta_{cd}+\delta_{\alpha\gamma}\delta_{ac}\delta_{\beta\delta}\delta_{bd}+\delta_{\alpha\delta}\delta_{ad}\delta_{\beta\gamma}\delta_{bc}\bigr)\nonumber\\
&-\frac{1}{2N(2N-1)(2N+2)}\bigl(\delta_{\alpha\beta}\delta_{ac}\delta_{\gamma\delta}\delta_{bd}+\delta_{\alpha\beta}\delta_{ad}\delta_{\gamma\delta}\delta_{bc}+\delta_{\alpha\gamma}\delta_{ab}\delta_{\beta\delta}\delta_{cd}+\delta_{\alpha\gamma}\delta_{ad}\delta_{\beta\delta}\delta_{bc}\nonumber\\
&\qquad\qquad\qquad\qquad\qquad\qquad+\delta_{\alpha\delta}\delta_{ab}\delta_{\beta\gamma}\delta_{cd}+\delta_{\alpha\delta}\delta_{ac}\delta_{\beta\gamma}\delta_{bd}\bigr).\label{fourO}
\end{align}
\end{widetext}
The average of Eq.\ \eqref{Grdef} becomes
\begin{equation}
\langle G\rangle=\tfrac{1}{4}G_{0}\left(2N+\frac{4Q^{2}-2N}{2N-1}\right),\label{Gaverage}
\end{equation}
which is just Eq.\ \eqref{eq:Gav}.

\end{document}